\documentclass{ws-ijmpcs}
\usepackage{amsmath,amssymb}
\usepackage{graphicx}
\usepackage{epsfig}
\usepackage{slashed}

\newcommand{\br}{{\bf r}}
\newcommand{\bu}{{\bf \delta}}
\newcommand{\bk}{{\bf k}}
\newcommand{\bq}{{\bf q}}
\def\({\left(}
\def\){\right)}
\def\[{\left[}
\def\]{\right]}

\begin{document}

\title{QUANTUM FIELD THEORY IN GRAPHENE\footnote{Based on a talk given by
D.~V.~Vassilevich at QFEXT 11, Benasque, September 2011.}}
\author{I. V. Fialkovsky}
\address{Instituto de F\'isica, Universidade de S\~ao Paulo,
Caixa Postal 66318 CEP 05314-970, S\~ao Paulo, S.P., Brazil\\
ifialk@gmail.com}

\author{D. V. Vassilevich}
\address{CMCC, Universidade Federal do ABC, Santo Andr\'e, S.P.,
Brazil\footnote{Also at Physics Department, St.Petersburg State University,
Russia}\\
dvassil@gmail.com}

\maketitle

\begin{abstract}
This is a short non-technical introduction to applications of the Quantum Field
Theory methods to graphene. We derive the Dirac model from the tight binding
model and describe calculations of the polarization operator (conductivity).
Later on, we use this quantity to describe the Quantum Hall Effect, light
absorption by graphene, the Faraday effect, and the Casimir interaction.
\end{abstract}

\ccode{PACS numbers: 73.63.-b, 11.10.Kk}

\section{Introduction}
Graphene, which is a one-atom thick layer of carbon atoms has many exceptional
properties\cite{gra-rev1,gra-rev2,gra-rev3} making it one of the most interesting
topics in condensed matter physics. The principle feature of graphene is that
the quasi-particle excitations satisfy the Dirac equation, where the speed of
light $c$ is replaced by the so-called Fermi velocity $v_F\simeq c/300$.
Therefore, the quantum field theory methods are very useful in the physics
of graphene. By applying these methods, one can explain anomalous Hall Effect
in graphene, the universal optical absorption rate, the Faraday effect, and
predict the Casimir interaction of graphene, and do much more.

The purpose of this article is to give a non-technical introduction to
and a short overview of the use of quantum field theory in graphene.
We start in the next Section with a derivation of the Dirac model from the
tight binding model and a discussion of possible generalizations of the former.
Quantum filed theory calculations in the Dirac model are presented in
Sec.\ \ref{sec-pol} at the example of polarization operator. This operator
is then used in Sec.\ \ref{sec-QHE} to explain the anomalous Hall conductivity
of graphene, which is proportional to $n+1/2$ with an integer $n$.
Next, in Sec.\ \ref{sec-abs}
we derive the universal absorption rate of $\alpha\pi \simeq 2.3\%$ for optical
frequencies. Sec.\ \ref{sec-Far} is devoted to explanations of the measured
giant Faraday effect in graphene and to further study of the polarization rotation.
Sec.\ \ref{sec-Cas} contains a survey of calculations of the Casimir
interaction of graphene.

\section{The Dirac model}\label{sec-Dir}
The Dirac model for quasi-particles in graphene was elaborated in full around 1984\cite{DIVM,Semenoff:1984dq} -- twenty years before actual discovery of graphene.
However, its basic properties, like the linearity of the spectrum, etc., were well known and
widely used much earlier due to the 1947 paper by Wallace\cite{Wallace}.
The purpose of most of the works of the time was to describe graphite rather than graphene.
For more details on the development and formulation of the Dirac model we refer
the reader to a recent review\cite{CastroNeto:2009zz}.

\begin{figure}
\centering \includegraphics[width=7cm]{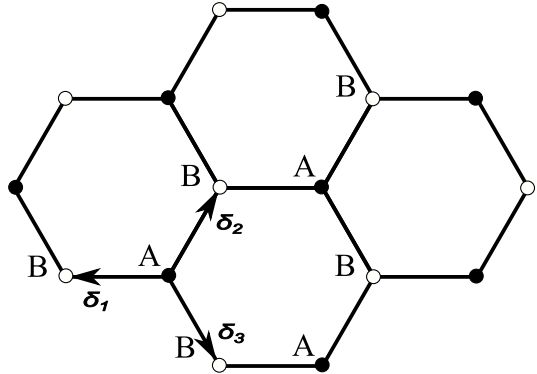} \caption{The honeycomb lattice
of graphene.} \label{lattice}
\end{figure}

In graphene, the carbon atoms form a honeycomb lattice (see Fig.\ \ref{lattice})
with two triangular sublattices A and B. The lattice spacing is $d=1.42$\AA.
The nearest neighbors of an atom from the sublattice A belong to the sublattice
B, and vice versa.
By adding to the location of an atom in the sublattice A any of the three vectors
\begin{equation}
\bu_1=d(-1,0),\qquad \bu_2 =d (1/2,\sqrt{3}/2),\qquad
\bu_3=d (1/2,-\sqrt{3}/2)\label{uuu}
\end{equation}
one arrives to the location of one of the nearest neighbors in the sublattice B.

In the \emph{tight binding model} only the interaction between electrons belonging to
the nearest neighbors
is taken into account, so that the Hamiltonian reads
\begin{equation}
H=-t\sum_{\alpha\in A}\sum_{j=1}^3
    \Bigl(a^\dag (\br_\alpha) b(\br_\alpha + \bu_j)
        +b^\dag (\br_\alpha +\bu_j)a(\br_\alpha)\Bigr)\,,\label{H1}
\end{equation}
where $t$ is the so-called hopping parameter, and the operators $a^\dag$, $a$, $b^\dag$,
$b$ are creation and annihilation operators of electrons in the sublattices A and B,
respectively. We adopt the units $\hbar = c =1$.
These operators satisfy usual anticommutation relations. Let us represent the wave function through
a Fourier transform.
\begin{equation}
|\psi\rangle =\Bigl( \psi_A(\bk) \sum_{\alpha\in A}e^{i\bk\br_\alpha} a^\dag(\br_\alpha)
    +\psi_B(\bk) \sum_{\beta\in B} e^{i\bk\br_\beta} b^\dag(\br_\beta) \Bigr) |0\rangle
\label{psi}
\end{equation}
The sublattice A is generated by shifts along the vectors $\bu_2-\bu_1$ and $\bu_3-\bu_1$.
Therefore, two momenta $\bk_1$ and $\bk_2$ are equivalent if
$(n_1(\bu_2-\bu_1)+n_2(\bu_3-\bu_1))\cdot (\bk_1-\bk_2)\in 2\pi \mathbb{Z}$ for all integer
$n_1$ and $n_2$. Representatives of the equivalence classes can be taken in a compact
region in the momentum space -- the Brillouin zone, which is a hexagon with the corners
at the points  $v_1=2\pi/3d\(1,1/\sqrt{3}\)$, $v_2=2\pi/3d\(1,-1/\sqrt{3}\)$,
$v_3=2\pi/3d\(0,-2/\sqrt{3}\)$, $v_4=-v_1$, $v_5=-v_2$ and $v_6=-v_3$. Opposite sides
of this hexagon are identified, and, in particular, the corners $v_1$, $v_3$ and $v_5$
are equivalent between themselves, as well as $v_2$, $v_4$ and $v_6$ are.

One can calculate
\begin{equation}
H|\psi\rangle =-t\Bigl( \psi_B\sum_{j=1}^3 e^{i\bk \bu_j}
 \sum_{\alpha\in A}e^{i\bk\br_\alpha} a^\dag(\br_\alpha)
+\psi_A \sum_{j=1}^3 e^{-i\bk \bu_j} \sum_{\beta\in B} e^{i\bk\br_\beta} b^\dag(\br_\beta) \Bigr) |0\rangle \label{Hpsi}
\end{equation}
so that the stationary Schr\"odinger equation $H|\psi\rangle =E|\psi\rangle$ becomes the
matrix equation
\begin{equation}
\left( \begin{array}{cc} 0 & -tX \\ -tX^* & 0 \end{array} \right)
\left( \begin{array}{c} \psi_A \\ \psi_B \end{array} \right)
=E \left( \begin{array}{c} \psi_A \\ \psi_B \end{array} \right)\,,\qquad
X=\sum_{j=1}^3 e^{i\bk \bu_j}\,. \label{mat}
\end{equation}
Clearly, the eigenvalues read
\begin{equation}
E=\pm t |X| \,.\label{EtX}
\end{equation}
The spectrum is symmetric, with positive and negative parts meeting at the
points where $X=0$. Solutions to this condition are easy to find, and they
coincide with the corners of the Brillouin zone. As we have discussed above, only
two of the corners are independent. Let us take $K_\pm =\pm v_6=\mp v_3$
as two independent solutions, called the Dirac points, describing two independent
ground states.

The next step is to expand the wave functions around these Dirac points,
$\psi^{\pm}_{A,B}(\bq )\equiv \psi_{A,B}(K_\pm +\bq)$. We suppose here
that $|\bq |$ is small compared to $1/d \sim 1{\rm KeV}$. Then,
one obtains the Hamiltonians
\begin{equation}
 H_\pm = \frac{3td}2 \left( \begin{array}{cc} 0 & iq_1\pm q_2 \\
-iq_1\pm q_2 & 0    \end{array} \right) =
v_F (-\sigma_2 q_1 \pm \sigma_1 q_2)\,,\label{Hpm}
\end{equation}
where $v_F$ is the Fermi velocity, $\sigma_i$ are the standard Pauli matrices. By substituting in $v_F=(3td)/2$ the
value of $d$ given above and $t=2.8{\rm eV}$ one obtains a number which
is slightly below the commonly accepted value $v_F\simeq 1/(300)$. By introducing
a four-component spinor $\psi\equiv (\psi_A^+,\psi_B^+,\psi_A^-,\psi_A^-)$,
we can unify $H_\pm$ in a single Dirac Hamiltonian $H=-i v_F\gamma^0 \gamma^a \partial_a$,
$a=1,2$, where we replaced the momenta $i\bq$ by partial derivatives, and
\begin{equation}
\gamma^0=\left( \begin{array}{cc} \sigma_3 & 0 \\ 0 & \sigma_3 \end{array}
\right),\qquad
\gamma^1=\left( \begin{array}{cc} i\sigma_1& 0 \\ 0 & i\sigma_1 \end{array}
\right),\qquad
\gamma^2=\left( \begin{array}{cc} i\sigma_2& 0 \\ 0 & -i\sigma_2 \end{array}
\right).\label{gam}
\end{equation}
These $4\times 4$ gamma matrices are taken in a reducible representation
which is a direct sum of two inequivalent $2\times 2$ representations.

Each of these two-component representations for graphene quasiparticles' wave-function
is somewhat similar to the spinor description of electrons in QED$_{3+1}$. However, in the case of graphene
this {\it pseudo}\,spin index refers to the sublattice degree of freedom rather than the real spin of the
electrons. The whole effect of the real spin, which by now did not appear in presented Dirac model,
is just in doubling the number of spinor components, so that we have 8-component
spinors in graphene ($N=4$ species of two--component fermions).

We conclude, that the tight binding model is equivalent below the energies
$1/d$ to a quasi-relativistic Dirac model, where the speed of light is replaced
by $v_F$. One should keep in mind that the tight binding model itself is an approximation.
One can improve this approximation by considering, for example,
couplings between atoms in the same lattice A or B, that is a
next-nearest neighbor coupling. The corresponding energy is estimated to be
about $0.1\, {\rm eV}$, much smaller than the nearest neighbor coupling $t$
considered above. Other possible couplings are summarized in Ref.\
\refcite{Gusynin:2007ix}. It is expected that with suitable modifications
the Dirac model is valid at least until the energies of $\sim 2\, {\rm eV}$.

Under various circumstances, it may make sense to consider the following
modifications of the Dirac model.
\begin{itemize}
\item One can add an interaction with the electromagnetic field. To preserve
gauge invariance, this interaction is introduced by replacing the usual partial derivatives
by gauge-covariant ones: $\partial \to \partial +ieA$. The electromagnetic
potential is not confined to the graphene surface, but rather propagates in the
ambient $3+1$ dimensional space. This field may be an external magnetic field, a classical
electromagnetic radiation, or quantized fluctuations.
\item
Quasiparticles may have a mass. This mass is usually very small, but the
introduction of a mass parameter may be convenient on theoretical grounds, e.g.,
to perform the Pauli-Villars regularization.
\item
A more important mass-like parameter is the chemical potential $\mu$, which
describes the quasiparticle density. This parameter can be easily varied in
experiments by applying a gate potential to graphene samples. Without a gate
potential, $\mu$ is usually very small in a suspended graphene, but can be
significant for epitaxial graphene due to interaction with the substrate.
\item
Impurities in graphene are described by adding a phenomenological parameter
$\Gamma$ which reminds an imaginary mass and enters the propagator of
quasiparticles exactly as $\epsilon$ in the Feynman prescription of contour
integration.
\item
Finally, most experiments with graphene are done at rather high temperatures,
which can be taken into account by usual rules of real or imaginary time
thermal field theory.
\end{itemize}

\section{Polarization operator}\label{sec-pol}
Let us proceed with quantum field theory calculations based on the
Dirac model.
From the Dirac Hamiltonian one can derive the action
\begin{equation}
S=\int d^3x \bar\psi \slashed{D}\psi \,,\qquad
\slashed{D}=i\tilde\gamma^j (\partial_j +ieA_j)+\dots \label{SDsl}
\end{equation}
where the dots denote any of additional terms described at the end of the previous
section. Tilde over $\gamma$ means that the space-components are rescaled
\begin{equation}
\tilde\gamma^0=\gamma^0,\qquad \tilde\gamma^{1,2}=v_F\gamma^{1,2} \,.\label{tilded}
\end{equation}
Quantized fermions give rise to an effective action for external electromagnetic
field (given by a sum of one-loop diagrams). To the second order in $A$, it reads
\begin{equation}
S_{\rm eff}(A)= A \ \raisebox{-3.75mm}
    {\psfig{figure=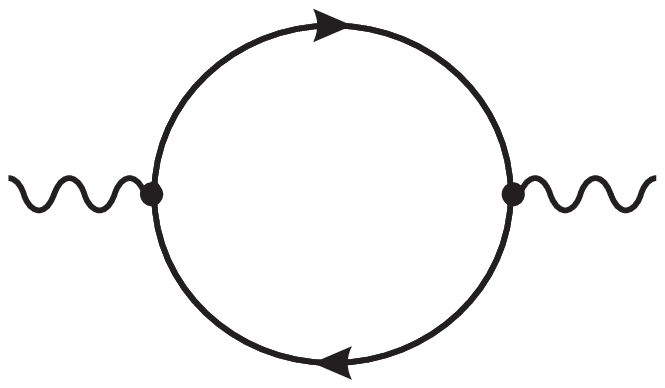,height=.4in}} \  A = \frac 12 \int
    \frac{d^3p}{(2\pi)^3} A_j(-p) \Pi^{jl}(p)A_l(p),\label{Seff}
\end{equation}
where $\Pi^{jl}$ is the polarization operator.

As an example, let us consider the polarization operator for a single
massive two-component fermion at zero temperature, zero chemical potential,
and without external magnetic field. Simple calculations give
\begin{equation}
    \Pi^{mn}
        = \frac \alpha{v_F^2} \, \eta^{m}_{j}\left[
            \Psi(p) \left(g^{jl}-\frac{\tilde p^j\tilde p^l}{\tilde p^2}\right)
            + i \phi(p) \epsilon^{jkl}\tilde p_k
            \right]\eta_l^n \label{Pmn}
\end{equation}
with $\epsilon^{jkl}$ being the Levi-Civita totally antisymmetric tensor normalized according
to $\epsilon^{012}=1$, $\eta_j^n={\rm diag}(1,v_F,v_F)$,
$\tilde p^m\equiv \eta_n^m p^n$. $\alpha = e^2/(4\pi) \simeq 1/137$ is the fine
structure constant.
The tensor structure of $\Pi^{mn}$ is uniquely
defined by quasi-relativistic and gauge invariances up to two functions,
$\Psi$ and $\phi$, which read
\begin{eqnarray}
&&\Psi(p)=
        \frac{2 m \tilde p -(\tilde p^2+4m^2){\rm arctanh }
({\tilde p}/{2m})}{2\tilde p},
    \label{Psi}\\
&&
\phi(p)
    =\frac{2m\, {\rm arctanh}(\tilde p/2 m )}{\tilde p}-1\label{phi}
\end{eqnarray}
here $\tilde p\equiv+\sqrt{\tilde p_j \tilde p^j}$, and we assume $m>0$.
From Eq.\ (\ref{Pmn}) one might get an impression that the polarization
tensor is singular at the $v_F\to 0$ limit, or, that $\Pi^{jk}$ is greatly
enhanced by the smallness of $v_F$. A more careful analysis shows that it
is not true, the polarization tensor remains finite in the $v_F\to 0$
limit provided the frequency is non-vanishing, $p_0\ne0$.

For more complicated external conditions, like a magnetic field, or for
a non-zero chemical potential, the quasi-Lorentz invariance is broken, and
the polarization tensor has a more complicated form than Eq.\ (\ref{Pmn}).

Particular importance of $\Pi^{jk}$ is due to the fact that this tensor
defines how the electromagnetic field propagates through graphene.
The full quadratic action for electromagnetic field is
$-\frac{1}4 \int d^4x\, F^2_{\mu\nu}+S_{\rm eff}$, where the effective action (\ref{Seff}) is
confined to the surface of graphene, which we place at $x^3=0$. The equations
of motion following from this action contain a singular term
\begin{equation}
\partial_\mu F^{\mu\nu} +\delta(x^3) \Pi^{\nu\rho}A_\rho =0.\label{Meq}
\end{equation}
We extended $\Pi$ to a $4\times 4$ matrix with $\Pi^{3\mu}=\Pi^{\mu 3}=0$.
The equations (\ref{Meq}) describe a free propagation of the electromagnetic field
outside the surface $x^3=0$ subject to the matching conditions
\begin{eqnarray}
&&A_\mu \vert_{x^3=+0}=A_\mu\vert_{x^3=-0},\nonumber\\
&&(\partial_3A_\mu)\vert_{x^3=+0}-
(\partial_3A_\mu)\vert_{x^3=-0}=\Pi_\mu^{\ \nu}A_\nu \vert_{x^3=0}
\label{match}
\end{eqnarray}
on that surface.

The polarization tensor and, more generally, Feynman diagrams involving
$2+1$ dimensional fermions were considered in a large number of papers.
Still in 1980's the function $\Psi$ (for $v_F=1$) was calculated in
Ref.\ \refcite{Appelquist}, while the pseudotensor part was discussed
about the same time in the context of the parity anomaly\cite{Sem,Red}.
A renormalization group approach to theories with $v_F\ne 1$ was
suggested in Ref.\ \refcite{Voz1}.
In this century, extensive calculations were done by the Kiev group
and collaborators\cite{Gusynin1,Gusynin2,Gusynin3,Pyatkovskiy}.
The formulas (\ref{Pmn}), (\ref{Psi}) and (\ref{phi}) are consistent with
that calculations.
\section{Physical effects}\label{sec-phy}
We proceed with considering some quantum field theory effects in graphene.
Specifically, we will be interested in applications of the polarization
tensor $\Pi^{ij}$. We shall see, that this quantity indeed defines important
and interesting physics.

\subsection{Quantum Hall Effect}\label{sec-QHE}
First of all, we note that the polarization tensor can be interpreted in terms of the conductivity of graphene.
Indeed, variation of the effective action (\ref{Seff}) with respect to $A_k$ produces
an expectation value of the electric
current in graphene, i.e. $j^k \simeq \Pi^{kl}A_l$. On the other hand, in the
temporal gauge, $A_0=0$, the electric field $E_a$ with the frequency $\omega$ is related
to the vector potential by $E_a =i\omega A_a$. By definition, the conductivity
is a matrix relating $j$ and $E$.
In this way, one arrives at the relation
\begin{equation}
\sigma_{ab}=\frac{\Pi_{ab}}{i\omega}\,, \quad a,b=1,2.\label{siPi}
\end{equation}

To study the conductivity tensor at zero frequency (the dc
conductivity) one puts the graphene sample in a constant magnetic field
perpendicular to its  surface and measures the anti-diagonal conductivity
as a function of the chemical potential $\mu$. This is precisely the set-up for
Hall experiments. For a one-layer graphene it was observed\cite{QHE1,QHE2} that the off-diagonal (Hall) conductivity is
quantized according to the law
\begin{equation}
\sigma_{12}\sim \Bigl( n+\frac 12\Bigr),\qquad n=0,1,2,\dots \label{QHE}
\end{equation}
i.e., the conductivity is proportional to half-integer numbers. This particular
type of the  Hall Effect is called anomalous, or unconventional integer,
or half-integer Quantum Hall Effect.  Starting with the Dirac model, the
behavior (\ref{QHE}) was predicted in Ref. \refcite{TQHE1} by using the
Feynman diagram approach described above, and in Ref.\
\refcite{TQHE2} from numerical simulations. This wonderful agreement between
theory and experiment was the first confirmation of existence of the Dirac
quasi-particles in graphene.

It is interesting to note, that the half-integer Quantum Hall Effect is observed
in the mono-layer graphene only. The double-layer graphene, for example,
exhibits conventional Integer Quantum Hall Effect. This does not however imply
that the Dirac model is not applicable to double-layer graphenes. The later case may be recovered if one carefully considers  phases
of determinant of the Dirac operator\cite{BS,BGSS}.

\subsection{Absorption of light}\label{sec-abs}
Another physical effect defined by the polarization operator is the absorption of light by
a suspended monolayer graphene.
In the simplest set-up one can neglect the chemical potential, suppose that there is no
external magnetic field and put temperature to zero, $T=0$. Consequently,
one can use the polarization operator (\ref{Pmn}), where, because
of $N=4$ generations of fermions in graphene, the scalar part $\Psi$
is multiplied by $N$, while the pseudo-scalar part $\phi$ cancells
out. The cancellation occurs due to the form of gamma-matices (\ref{gam}),
containing two inequivalent representations related by the parity
transformations. In other words, we need to make a substitution
\begin{equation}
\Psi \to \Psi_N=N\Psi,\qquad \phi \to 0\,.\label{subs}
\end{equation}
Let us consider a plane wave with the frequency $\omega$
propagating along the $x^3$-axis
from $x^3=-\infty$ with the initial polarization parallel to $x^1$,
which is being reflected by and transmitted through
the graphene sample
\begin{equation}
    A=e^{-i\omega t}\left\{\begin{array}{ll}
        \mathrm{\bf e}_x e^{ik_3 x^3}
            + (r_{xx}\mathrm{\bf e}_x
          +r_{xy}\mathrm{\bf e}_y)e^{-ik_3 z}, & x^3<0 \\
        (t_{xx}\mathrm{\bf e}_x +t_{xy}\mathrm{\bf e}_y)e^{ik_3 z}, & x^3>0 \\
                \end{array}\right.\label{az}
\end{equation}
where $\mathrm{\bf e}_{x,y}$ are unit vectors in the direction $x^{1,2}$.
The mass-shell condition (free Maxwell equations away of the graphene sample) implies $k_3=\omega$.
For such waves the matching conditions (\ref{match}) simplify,
\begin{eqnarray}
&& A_a\big|_{x^3=+0}=A_a\big|_{x^3=-0} \nonumber\\
&&  (\partial_z A_a)_{x^3=+0}-(\partial_z A_a)_{x^3=-0}
=\alpha \Psi_N (k) \delta_a^b \,,\label{mc2}
\end{eqnarray}
where we used (\ref{Pmn}) and (\ref{subs}). The transmission coefficients
can be easily found, see e.g. Ref.\ \refcite{Fialkovsky:2009wm},
\begin{equation}
t_{xx}=\frac{2 \omega}{ i\alpha \Psi_N+2\omega}\,,\qquad
t_{xy}=0 \,.\label{tt}
\end{equation}
The intensity of transmitted light thus reads
\begin{equation}
\mathcal{I}=|t_{xx}|^2=1 + \frac{\alpha{\mathrm{Im}}\, \Psi_N}{\omega}
+O(\alpha^2) \,.\label{inten}
\end{equation}
One can reformulate this result in terms of conductivity by noting that $\alpha\Psi= i\omega\sigma_{xx}$, as follows from (\ref{Pmn}) and (\ref{siPi}).

At large frequencies, $\omega \gg 2m$, we have $\Psi \simeq -i\pi\omega/4$,
yielding $\alpha {\mathrm{Im}}\, \Psi_N/\omega\simeq -\alpha\pi$.
The same conclusion is also valid if the polarization tensor is calculated
with more general external conditions\cite{GSC}.
Therefore, we confirm the prediction of Refs.
\refcite{ando,Fal,stauber,abs2} made on somewhat different theoretical grounds
of universal absorption rate of $\alpha\pi \simeq 2.3$\%, that was
confirmed by the experiment\cite{abs1}.

Clearly, this uniform absorption rate is much larger than one would expect from
a one-atom thick layer.

\subsection{The Faraday effect}\label{sec-Far}
The Faraday effect reminds very much the Hall effect at ``non-zero frequencies'',
and this analogy was used in Ref.\ \refcite{VoMi} to conjecture that the
former should be common for Hall systems\footnote{In Ref.\
\refcite{Fialkovsky:2009wm} the Faraday rotation was related to possible
non-compensation of parity-odd parts of the polarization tensor between various generations
of fermions.}.
The set-up is, therefore, very similar: a graphene sample subject to a
constant magnetic field perpendicular to its surface. Instead of the Hall conductivity,
we shall be interested in the rotation of the polarization plane of a light beam passing
through the surface of graphene.
The frequency of photons shall be kept as a variable parameter, as well as the chemical
potential. It can be shown that it is usually sufficient to consider the zero-temperature case only,
but impurities are essential.

By solving again the matching conditions (\ref{match}) for plane wave (\ref{az}), but now 
with a polarization operator calculated in presence of constant magnetic field, one finds that the angle
$\theta$ of polarization rotation and the intensity $\mathcal{I}$ of transmitted light are given by
\begin{equation}
    \theta= -\frac{{\rm Re}\sigma_{xy}}2+O(\alpha^2), \qquad
        \mathcal{I}= 1 - { {\rm Re}\sigma_{xx}}+O(\alpha^2)\,,
    \label{Tth_a}
\end{equation}
where we used (\ref{siPi}) to express the $\Pi$ components through diagonal and Hall conductivities of graphene.

An experiment\cite{Fara} made recently demonstrated a ``giant'' Faraday rotation angle
of about $0.1$rad peaked at low frequencies for a magnetic field of about $7$~Tesla.
A theoretical study\cite{FV} shows a good agreement between this experiment
and the Dirac model. Besides, the Dirac model predicts other effects, like
step-function like behavior of the rotation angle and the peaks at higher
frequencies. Another interesting theoretical observation is that although
the data of Ref.\ \cite{Fara} can be nicely fitted by the Drude formula
for conductivity, this Drude-like behavior cannot be uniformly extended
for all frequencies.

\subsection{The Casimir effect}\label{sec-Cas}
The Casimir effect\cite{Bordag:2001qi}, which is one of the main topics of this
Workshop, is sometimes understood as any manifestation of the zero point energy.
We consider the Casimir effect in a stricter sense, as an interaction of two
uncharged well-separated bodies due to quantum fluctuations of the electromagnetic
vacuum. In the framework of the present review, let us take a suspended graphene sample separated by the
distance $a$ from a parallel plane ideal conductor. Under these conditions,
one can neglect $m$, $\mu$, and $\Gamma$, but the temperature will be non-zero,
in general.

The lowest-order diagram which gives the Casimir free energy is\cite{Bordag:2009fz}
\begin{equation}
\mathcal{F}_1\sim \ \raisebox{-3.75mm}
    {\psfig{figure=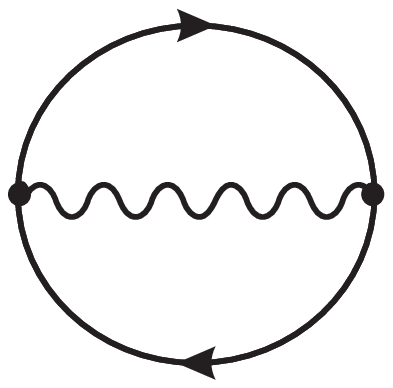,height=.4in}} \ , \label{E1}
\end{equation}
where the photon propagator satisfies conductor boundary conditions on the surface
$x^3=a$. (We use the free energy since this is the relevant quantity at
finite temperature).
In this expression, one of the boundaries (conductor) is taken into account
exactly, while the other one (graphene) - perturbatively, at the first order of
$\alpha$. A better approximation may be obtained by considering a closed loop of a propagator
satisfying both conductor boundary conditions at $x^3=a$ and the matching conditions
(\ref{match}) at $x^3=0$. This boils down to the use of the Lifshitz\cite{Lifshitz}
 formula
\begin{equation}
    {\mathcal F}
    =T\sum_{n=-\infty}^\infty\int\frac{d^2{\bf p}}{8\pi^2} \ln [(1-e^{-2p_\| a}r_{\rm
 \rm TE}^{(1)}r_{\rm \rm TE}^{(2)})
        (1-e^{-2p_\| a}r_{\rm \rm TM}^{(1)}r_{\rm \rm TM}^{(2)})] \,,
        \label{EL}
\end{equation}
where $p_\|=\sqrt{\omega_n^2+{\bf p}^2}$, and $\omega_n=2\pi n T$ are the
Matsubara frequencies.
$r^{(1,2)}_{\rm TE,TM}$ are the reflection coefficients for the TE and TM modes at
each of the two surfaces. For the second surface, which is an ideal conductor,
we have $r_{\rm \rm TM}^{(2)}=1$, $r_{\rm \rm TE}^{(2)}=-1$. The reflection
coefficients for graphene are calculated similarly to the case considered
in Sec.\ \ref{sec-abs}, though at non-zero temperature and arbitrary tangential
momenta the calculations are more complicated.

It is an interesting exercise to check that at the $\alpha^1$ order the Lifshitz
formula reproduces the two-loop diagram (\ref{E1}). At zero temperature both
approaches give consistent result of the order of $2.7\%$ (for the Lifshitz
formula) of the Casimir interaction between two ideal metals\cite{Bordag:2009fz}
In the view of Sec.\ \ref{sec-abs}, this is not very surprising.  Some unexpected
features appear at non-zero temperature\cite{Fialkovsky:2011pu}. The perturbative
result (\ref{E1}) rapidly becomes unreasonably large for growing $T$ signalling that we have left
the perturbative region. Roughly speaking, this effect is caused by a competition
between two small parameters, $\alpha$ and $v_F$.
The non-perturbative free energy (\ref{EL}) also grows
with increase of the dimensionless parameter $aT$, and, at the large $T$
asymptotics we have
\begin{equation}
{\mathcal{F}}\vert_{T\to \infty} \simeq -\frac{T\zeta(3)}{16\pi a^2} \,,
\label{Fas}
\end{equation}
which is just a half of the interaction between two ideal metals in the same
regime, or is the same value as for non-ideal metals described by the
so-called Drude model! Thus, the Casimir interaction of graphene at high
temperature is extremely strong. This agrees qualitatively with
Ref.\ \refcite{Gomez} where the Casimir interaction of two graphene samples
was considered.

We conclude this Section with some references. Other papers which study
the Casimir effect for graphene are Refs.\ \refcite{Dobson,Sernelius,Ali}.
Some earlier calculations used the hydrodynamic model for the electrons in
graphene\cite{h1,h2}. Since this model does not reproduce the linear
dispersion law characteristic for graphene, this line of research was abandoned.
More details on the presents status of Casimir effect in graphene can be found
in Ref.\ \refcite{Mar}.

\section{Conclusions}
The main message of this paper is that quantum field theory calculations based on the Dirac
model of quasiparticles are extremely effective in describing the physics of
graphene. One of
the reasons for this effectiveness is the equivalence between the tight binding model and
the Dirac model for small momenta. It is interesting to note that a single one-loop
diagram of the polarization tensor considered in Sec.\ \ref{sec-pol} is responsible for
many physical phenomena, such as the Hall and Faraday effects and the uniform light absorption
rate (where all experiments are in a good agreement with theory), and the Casimir interaction
of graphene (where no experiment has been done so far). All the effects discussed
above are very strong, much stronger than one would expect from a one-atom
thick layers.

Since parameters of the Dirac model may differ considerably from sample to sample of
graphene, it makes sense to perform various types of experiments with the same samples.
E.g., one can combine optical measurements with Casimir experiments.

Some topics were not considered here, though they definitely deserve being mentioned.
One of such topics is the graphene nanoribbons. Before calculating the polarization
tensor, one should define boundary conditions which are compatible with quantum field
theory. Such an analysis was performed in Ref.\ \refcite{BSrib}.
Another extremely intersting topic is the topological effects in graphene (see
Refs.\ \refcite{topol,Voz2} for a review), which includes the Jackiw-Pi model\cite{JPi},
applications of the index theorem, curvature effects, etc. This list of missing points
is not exhaustive. There is much more in the area of applications of Quantum Field
Theory to graphene.

\section*{Acknowledgments}
We are grateful to M.~Bordag, D.~Gitman and V.~Marachevsky for collaboration,
to G.~Beneventano and M.~Santangelo for fruitful discussions, and to the Organizers
of QFEXT 11 for making this enjoyable workshop and support. This work was supported
in parts by FAPESP (I.V.F. and D.V.V.) and by CNPq (D.V.V.).

\end{document}